\documentclass[a4paper, 12pt]{article}
\newcommand{\n}{
\noindent
{\scriptsize Proc.Int.Conf. "Geometrization of Physics III"\\
Kazan State University, Kazan, October 1-5, 1997}
}
\begin{document}
\n
\begin{center}
{\large \bf Particles as singularities within the unified\\  
algebraic field dynamics}

\bigskip
V.V.Kassandrov, J.A.Rizcalla

\medskip
{\it Dept.Gen.Phys.\\
     Ordjonikidze Str. 3\\
     People's Friendship University\\
     117419 Moscow Russia\\
     tel.: (095)1387750\\
     e-mail: vkassan@mx.pfu.edu.ru}
\end{center}

\bigskip
{{\bf Abstract.} We consider a nonlinear generalization of Cauchy-Riemann eqs. 
to the algebra of biquaternions. From here we come to "universal generating 
equations" (1) which deal with 2-spinor and gauge fields and form the basis 
of some unified algebraic field theory. For each solution of eqs.(1) 
the components of spinor field satisfy the eikonal and d'Alembert eqs., 
and the strengths of gauge field - both Maxwell and Yang-Mills eqs. We 
reduce eqs.(1) to that of shear-free geodesic null congruence and
integrate them in twistor variables. Particles are treated as concurrent 
singularities of the effective metric and the electromagnetic field. For 
unisingular solutions the electric charge is quantized, and the metric is of 
Schwarzschild or Kerr type. Bisingular solutions are announced too.}

\bigskip
\noindent
{\bf1.Algebrodymamical approach to field theory and universal generating 
equations}

In the general framework of {\it algebrodynamical} paradigm (see 
~\cite{AlgDyn, CCF, Klu} and references therein) it was proposed to regard 
the set of equations
$$
d\xi = A(x)*dX*\xi(x)
\eqno(1)
$$
as the only basis of some unified non-Lagrangian field theory. In formula (1) 
the asterisk denotes multiplication in {\it the algebra of biquaternions} $B$ 
(equivalent to multiplication of matrices), and $X$ represents the $2\times2$ 
Hermitian matrix of space-time coordinates. The two-column variables $\xi(x)$ 
may be identified as a fundamental spinor field (related to a null complex 
tetrad, see Section 6), while the components $A_\mu(x)$ of the $2\times2$ 
matrix $A=A^\mu(x)\sigma_\mu$ be considered as the $C$-valued electromagnetic 
(EM) potentials.

The properties and interpretation of the eqs.(1) are only examined throughout 
this article. They originate from the {\it $B$-generalized Cauchy-Riemann 
equations} (Section 2), appear to be Lorentz and gauge invariant (Section 3) 
and impose strict limitations on both the spinor and EM fields (Section 4). 
Indeed, for each solution to the eqs.(1) {\bf the components of spinor 
field satisfy the eikonal and d'Alembert eqs. (Sections 2,5), while the EM 
field strengths obey the Maxwell eqs. for free space}. Moreover, a close 
connection exists between the solutions to the eqs.(1) and those to the 
vacuum Yang-Mills and Einstein eqs. (Sections 4 and 6 respectively).

In view of the above relations between the wave-like, gauge and GTR equations 
(we'll call them {\it conventional equations} (CEqs) for brevity) the eqs.(1) 
have the right to be called {\it universal albebraic field equations} or, 
briefly, universal eqs. (UEqs). Since the CEqs are all of vacuum type, in the 
approach we develop {\bf the particles may be regarded as nothing but 
{\it singularities} of the fields}. We'll see (Sections 5,7) that the 
structure of singularities of the CEqs (including even the free Maxwell ones) 
is surprisingly rich, complicated (point-, string- or toroidal-like) and 
possibly unknown up to now.

On the other hand, the characteristics and the time evolution of the 
singularities are completely governed by the {\it over-defined} nonlinear 
structure of UEqs (1), since {\bf the CEqs are only necessary not sufficient 
{\it compatibality conditions} in respect to the primary system of UEqs}. In 
this way (Section 5), the Coulomb Ansatz corresponds to some solution of 
UEqs iff {\bf the value of electric charge of the source is fixed to be unit}, 
this in spite of linearity of Maxwell eqs. themselves. Thus, the quantization 
holds here just on a classical level and due again to a strict {\it 
over-defined} structure of UEqs (1). 

From the other results first presented, the relation between the UEqs and the 
system of {\it shear-free geodesic null congruence} from GTR should be 
distinguished. In its account, the integration of the system (1) is fulfilled 
in {\it twistor variables} (Section 6). On the other hand, this permits to 
define an effective {\it Riemann metrics} for each solution to UEqs. In the 
stationary axial-symmetric case, these metrics appear to satisfy the vacuum 
Einstein eqs. and are just of Schwarzschild or Kerr type!

In Section 7 we discuss general interpretation of particles as singularities, 
and bisingular solution to the UEqs is announced in this context.

\bigskip
\noindent
{\bf 2.Algebraic origination and 2-spinor structure of universal equations}

Let $A$ be a finite-dimensional {\it associative} and {\it commutative} 
algebra over $R$ or $C$. The natural definition of $A$-differentiability  was 
proposed by G.Sheffers as far as in 1893 and has the form 
(see ~\cite{Shir}, chapter 4 for details):
$$
dF = D(Z) * dZ ,
\eqno(2)
$$
($*$) being multiplication in $A$, $F(Z)$ being an $A$-valued function 
of $A$-variable $Z \in A$, and $D(Z) \in A$ - some other $A$-function as well.

The eqs.(2) may be considered as {\it the condition of $A$-valued differential 
$1$-form to be exact}\footnote{The usual conditions of smoothness of $F(Z)$ -
components and existence of a positive norm in $A$-space are supposed to  
be fulfilled}. 
For a particular case of complex algebra $A \equiv C$ the eqs.(2), after 
elimination of the components of $D(Z)$, lead to the Cauchy-Riemann 
(CR) equations of ordinary form. 

To succeed in the formulation of differentiability conditions for the case of 
associative {\it noncommutative} algebra $G$ one notices that the most general 
component-free form of infinite- simally small increment of a $G$-function is 
\footnote{For example, in the simplest case of a quadratic function $F(Z) = 
Z*Z$ one has
$$
dF = Z*dZ*E + E*dZ*Z,
$$
$E$ being the unit element in $G$.}

$$
dF = L_1(Z) * dZ * R_1(Z) + L_2(Z) * dZ * R_2(Z) + ... ,
\eqno(3)
$$
the set of pairs $\{L_i(Z), R_i(Z)\}$ replacing the derivative $D(Z)$ of the 
commutative case. Notice that just the representation (3) serves infact as a 
basis of the noncommutative analysis in the version recently presented 
~\cite[chapter 7]{Chrenn}.

Unfortunately, {\bf no relation is known to exist, generally, between the 
components of a "good" $G$-function}, i.e. a function whose differential 
may be presented in the form (3). This is quite contrary to the situation 
in the commutative case, in the $C$-case with CR-equations in particular.
Besides, from geometrical point of view, the functions satisfying to eqs.(3) 
show no analogy to the conformal mappings of the complex analysis. For these 
reasons the version of A.Yu.Chrennikov cannot be regarded as fully 
successful.

The direct account of noncommutativity in the very definition of 
$G$-differentiability seems, however, quite natural and promising. In 1980 just this way 
towards the construction of noncommutative analysis was proposed by one of 
the authors ~\cite{Depon} (see ~\cite{AlgDyn} and the references therein). 
On the other hand, in order to have some restrictions on the components of 
$F(Z)$ ({\it generalized CR-equations}) it was proposed to regard as "true" 
$G$-differentiable only such $G$-functions for which {\bf representation (3) 
is reduced to one "elementary" $G$-valued differential $1$-form only}, i.e. 
for which it holds
$$
dF = L(Z) * dZ * R(Z) ,
\eqno(4)
$$
$L(Z), R(Z) \in G$ being called {\it semi-derivatives} of $F(Z)$ (they are 
defined up to an element from the {\it centre} of $G$, see ~\cite{AlgDyn}).

Definition of $G$-differentiability (4) may be considered as requirement 
on an elementary $G$-valued  $1$-form to be exact \footnote{Note 
that the {\it elementary} $G$-form may be defined as {\it the most general 
$G$-valued $1$-form} which may be constructed by means of operation of 
multiplication in $G$ only}. For $G$ being commutative again, the conditions 
(4) are evidently reduced to the old ones (2) (and so to the CR-equations in 
$C$-case).

The definition (4) appreciably narrows down the class of "good" $G$-functions, 
cutting off, say, all the polynomials (exept the trivial linear one). This 
certainly seems rather unexpected from the standpoint of customary complex 
analysis. Nevertheless, it singles out just such a class of $G$-functions 
which is natural by algebraic considerations, extremely interesting in geometrical 
properties and admits a wonderful field-theoretical interpretation.

In the exclusive case of real Hamilton quaternions $G \equiv H$ eqs.(4) 
appear to be just {\bf the algebraic condition for the mapping $F: Z 
\rightarrow F(Z)$ to be {\it conformal} in $E^4$} (see ~\cite{AlgDyn},~\cite{Klu} 
for details). However, since the conformal group of $E^4$ is known to be finite (15-) 
parametric, the functions corresponding to eqs.(4) are rather trivial to be 
treated, say, as field variables. Fortunately, the situation becomes quite 
different when one turns {\bf to the complex extention of $H$, i.e. to the 
algebra of {\it biquaternions} $B$}, which only we are going to deal with 
below.\footnote{Some considerations about differentiability in Dirac-Clifford 
and octonion algebras are presented in ~\cite[chapter~1]{AlgDyn}}

For $B$-algebra the $2\times2$ complex matrix representation is 
suitable. For it we take
$$
Z \Leftrightarrow {z^0 + z^3~~~z^1 - i z^2 \choose z^1 + i 
z^2~~~z^0 - z^3} \equiv {u ~~ \bar w \choose w ~~ v} \equiv z^\mu \sigma_\mu ,
\eqno(5)
$$
where $z^\mu \in C$ and $\sigma_\mu = \{E,~ \sigma_a\}$ being the unit and 
three 
Pauli matrices respectively (as usual, $\mu,\nu,...=0,1,2,3$ and 
$a,b,...=1,2,3$), 
and $u,v = z^0 \pm z^3;~ w,\bar w = z^1 \pm i z^2$ being spinor 
coordinates to be used below). 
Applying now the column or the full row-column splitting to the eqs.(4) we 
come to the following two forms:
$$
d\xi  = L(Z) * dZ * \eta (Z), \qquad df = \phi (Z) * dZ * \psi (Z) ,
\eqno(6a,b)
$$
$\xi (Z), \eta (Z) \in C^2$ being two 2-columns and $\phi (Z), \psi (Z) \in 
C^2$ - two 2-rows, mutually independent in general, $f(Z) \in C$ - some any matrix 
component of $F(Z)$. According to the symmetry properties of the eqs.(6a,b) 
the quantities $\xi ,\eta ,\phi ,\psi$ manifest themselves as 2-spinors, 
whereas $f(Z)$ as a scalar (see Section 3 and the article 
~\cite{GravCosm} for details).

From the condition (6b) in account of well-known {\it Fiertz identities}, 
{\bf the (complexified) eikonal equation for each (matrix) component $f(Z)$ of 
a $B$-differentiable function $F(Z)$ now follows} ~\cite{Westnik, AlgDyn}
$$
\eta^{\nu\lambda} \partial_\nu f \partial_\lambda f = 0,
\eqno(7)
$$
$\partial_\nu \equiv \partial / \partial z^\nu$ being the partial derivatives 
and $\eta^{\nu\lambda}$ being the metric tensor, of the form 
$\eta^{\nu\lambda} = diag(+1,-1,-1,-1).$ in representation (5).

The eikonal equation (7) plays in $B$-analysis the role similar to that of the 
Laplace equation in two-dimensional complex case. Thus, the definition (4) of 
$G$-differentiability links together the {\it noncommutativity} of $G$-algebra 
(figurating directly in (4)) with the {\it nonlinearity} of the generalized 
CR-equations consequent. There is nothing surprising in this correlation from 
the usual standpoint of gauge theories (non-Abelian group results in a 
nonlinear 
structure of Yang-Mills strengths). However, within the framework of the 
noncommutative analisis this interrelation was proposed and demonstrated, 
perhaps, for the first time (all of the previous works  
deal with trivial linear generalizations of the CR-equations, see for example 
~\cite{Shir, Gursey}).

Noticing that the eqs.(6b) follow directly from the eqs.(6), and the 
latters make it possible to reconstruct an arbitrary solution to the entire 
system (4), we come to the {\it fundamental 2-spinor structure} for the 
primary 
system (4) to possess. Together with its nonlinear character this allows us to 
formulate a field theory on the base of eqs.(6a) only. We hope to do this 
elsewhere, whereas here we restrict ourselves with a particular case $\xi(Z) 
\equiv \eta(Z)$ in eqs.(6a).  

Aa the only {\it ad hoc} conjecture we are obliged to take here 
is the requirement for coordinates $z^\mu$ in 
(5) to be real, $z^\mu \equiv x^\mu \in R$, i.e. to belong to the 
{\it Minkowsky space} being a subspace of the entire complex 
vector space of $B$-algebra (this requires for the coordinate-representing 
matrices in (5) to be Hermitian, $Z \equiv X = X^+$).

In account of these two limitations, the second one  clearly necessary to 
ensure the 
{\it relativistic invariance} of theory, we come back to the UEqs (1), 
announced at 
the beginning, as the equations of some {\bf algebraic nonlinear field 
theory}, 
having something to do with a spinor field as well as with gauge 
fields 
(see next Section). It is obvious, however, that such a theory will 
be very exotic due to the {\bf over-defined, nonLagrangian structure} of its 
only dynamical background - the UEqs, to which we now proceed.

\bigskip
\noindent
{\bf 3.Geometrodynamical interpretation and gauge structure of UEqs.}

In a 4-index form the UEqs (1) take the form
$$
\partial_\nu \xi = A(x)*\sigma_\nu *\xi(x),
\eqno(8)
$$
where $A(x) \equiv L(X) = A^\mu (x)\sigma_\mu$. According to (1) or (8), the 
UEqs. may 
be considered in the framework of geometrodynamics as {\bf the conditions for 
a fundamental spinor field $\xi(x)$ to be covariantly constant} in respect to 
the effective affine connection
$$
\Gamma = A(X)*dX, \qquad \Gamma_\nu = A(x)*\sigma_\nu
\eqno(9)
$$
which may be called {\it left $B$-connection}. It is determined by the 
structure 
of $B$-algebra and induces a specific affine geometry of {\it Weyl-Cartan 
type} 
on the complex vector space of $B$-algebra. To see this, one should return 
back 
in (8) from the spinor $\xi(X)$ to a full $2\times2$ matrix $F(X) = F^\mu (x) 
\sigma_\mu$; then one gets for the components 
$$
\partial_\nu F^\mu = \Gamma^\mu_{\nu\rho} (A(x)) F^\rho(x),
\eqno(10)
$$
where the connection coefficients have a specific form
$$
\Gamma^\mu_{\nu\rho} (A) = \delta^\mu_\nu A_\rho + \delta^\mu_\rho A_\nu - 
\eta_{\nu\rho} A^\mu - i \varepsilon^\mu_{.\nu\rho\lambda} A^\lambda,
\eqno(11)
$$
including the {\it Weyl nonmetricity} and the {\it 
torsion} 
terms related to each other (the Weyl's vector $A_\mu(x)$ being proportional 
to the pseudo-trace of the torsion tensor $i A_\mu(x)$).

Note that the $B$-induced complex Weyl-Cartan connection (11) was proposed 
firstly in ~\cite{AlgDyn} and recently used by V.G.Kretchet in his search 
for 
geometric theory of electroweak interactions ~\cite{Kretch} (based on the 
break of $P$-invariance by the torsion term in (11)).

The UEqs (1) are evidently form-invariant under the global transformations 
of coordinates and field variables
$$
X \Rightarrow X' = M^+ * X * M, \qquad 
\xi \Rightarrow \xi' = M^{-1}\xi, \quad A \Rightarrow A' = 
M^{-1}*A*(M^+)^{-1},
\eqno(12a,b)
$$
$M \in SL(2,C)$ being an arbitrary unimodular ($\det \Vert M \Vert =1$) 
$2 \times 2$ complex matrix.

The 6-parametric group of transformations (12a)~ $2:1$~ corresponds to the 
continious transformations of the coordinates $x^\mu$ ($X=x^\mu \sigma_\mu$) 
from Lorentz group. Thus, the  {\bf UEqs. are relativistic invariant} and, 
according to the laws of transformations (12b), the quantities $A_\mu(x)$ and 
$\xi_B (x), B=1,2$ behave themselves as the components of a 4-vector and a 
2-spinor respectively.
As to the local symmetries of UEqs., the system (8) may be showm to preserve 
its form under so called "{\it weak gauge transformations}" ~\cite{GravCosm}
$$
\xi_B \Rightarrow \xi'_B = \lambda \xi_B, \qquad A_\mu \Rightarrow A'_\mu = 
A_\mu + \frac{1}{2} \partial_\mu \ln \lambda,
\eqno(13)
$$
$\lambda \equiv \lambda (\xi^1,\xi^2) \in C$ being some rather smooth scalar 
function {\bf dependent on two spinor components of the starting solution 
only} (instead of its direct dependence on all 4-coordinates $x^\mu$ in a 
usual 
gauge approach)!\footnote{Besides, the UEqs. are invariant under the gauge 
transformations of Weyl type, related to the conformal change of metric; a 
discussion of the {\it double gauge group} so arising may be found in 
~\cite{CCF}.}

In account of the gauge nature of the 4-vector $A_\mu(x)$ and of its close 
relation to the Weyl's nonmetricity vector, it seems quite natural {\bf to 
identify $A_\mu(x)$} (up to a dimensional factor) {\bf with the 4-vector of 
potentials of a ($C$-valued) electromagnetic field}. Leaving for the next 
Section the discussion of complex structure of EM-field, we recall only that 
both the spinor and EM fields may be obtained from the only system (8) due to 
its over-defined structure. Then the problem arises what sort of restrictions 
on EM-strengths are imposed by UEqs, and in what way are they related to 
Maxwell equations?

\bigskip
\noindent
{\bf 4.Self-duality, Maxwell \& Yang-Mills equations as the consequences of 
UEqs.}

Since the set of UEqs (1) or (8) is over-defined (8 eqs for 2 spinor plus 
4 potential components), some {\it compatibility conditions} should be 
satisfied. In particular, commutators of partial derivatives $\partial_
{[\mu} \partial_{\nu ]} \xi = 0$ in (8) should turn to zero, this 
being correspondent to the {\it closeness} of the $B$-valued 1-form in 
(1) owing to the Poincar\'e lemma. After the calculations being derived we get 
$$
0 = R_{[\mu\nu]} \xi ,
\eqno(14)
$$
where the quantities 
$$
R_{[\mu\nu]} = \partial_{[\mu} A\sigma_{\nu]} - [A\sigma_\mu ,~A\sigma_\nu]
\eqno(15)
$$
represent the {\it $B$-curvature tensor} of the $B$-connection (9).

From (14) it doesn't follow $R_{[\mu\nu]} \equiv 0$, since the 
spinor $\xi(x)$ is not an arbitrary one. However, it may be shown (see ~\cite
{AlgDyn},~\cite{GravCosm}) that the {\it self-dual part} $R^+_{[\mu\nu]}$ 
of (15) 
$$
R^+_{[\mu\nu]} \equiv R_{[\mu\nu]} + \frac{i}{2} \varepsilon_{\mu\nu}^{..\rho
\lambda} R_{[\rho\lambda]} =0
\eqno(16)
$$
should turn to zero as a consequence of eqs.(14). Being written in components, 
the expressions (15),(16) result in 3+1 equations 
$$
{\cal F}_{[\mu\nu]} + \frac{i}{2} \varepsilon_{\mu\nu}^{..\rho\lambda}
{\cal F}_{[\rho\lambda]} = 0 ,
\eqno(17)
$$
$$
\partial_\mu A^\mu + 2 A_\mu A^\mu = 0 ,
\eqno(18)
$$
where the tensor 
$$
{\cal F}_{[\mu\nu]} = \partial_{[\mu}A_{\nu]}
\eqno(19)
$$
is a usual strengths' tensor of EM field. 3-vector form of the eqs.(17) 
$$
\vec{\cal E} + i\vec{\cal H} = 0
\eqno(20)
$$
relates the ($C$-valued) {\it electric} $\vec{\cal E}$ and {\it magnetic} 
$\vec {\cal H}$ vectors of field strength tensor
$$
{\cal E}_a = {\cal F}_{[oa]} = \partial_o A_a - \partial_a A_o , ~~~ {\cal 
H}_a = \frac{1}{2} \varepsilon_{abc} {\cal F}_{[bc]} = \varepsilon_{abc}
\partial_b A_c .
\eqno(21)
$$
Thus, we see that {\bf the self-duality conditions (17) and the 
"inhomogeneous Lorentz condition" (18) \footnote{Geometrically it corresponds 
to the condition for the scalar 4-curvature invariant of the Weyl tensor to be 
null, see ~\cite{GravCosm}} appear to be the integrability conditions of 
the UEqs.}

According to the definitions of field strengths through the potentials (21) 
and 
to the self-duality conditions (20), we conclude then that {\bf the free-space 
Maxwell equations are satisfied identically for each solution to the UEqs!} 

The complex nature of field strengths (21), however, doesn't manifest itself 
in doubling of the degrees of freedom' number of EM field owing just to the 
self-duality eqs.(20); from the latters we get only 
$$
\vec B = \vec E , ~~~~~~~\vec D = -\vec H ,
\eqno(22)
$$
$\{\vec E, \vec H\}$ and $\{\vec D, \vec B\}$ being respectively the real ($\Re$) 
and imaginary ($\Im$) parts of the primary complex fields $\{\vec{\cal E},
\vec{\cal H}\}$. The real-part fields $\vec E$ and $\vec H$ are therewith 
mutually independent in algebraic sense and satisfy the Maxwell eqs. owing to 
linearity of the latters. \footnote{The same being true, of course, for the $\Im$-part 
fields $\vec D, \vec B$,~ providing in account of (22) a dually-conjugate 
solution to the Maxwell eqs.}. 

The meaning of decomposition of the unique complex field into the real and 
imaginary parts is that the density of conservative {\it energy-momentum 
tensor} can be defined through the latters in a usual way, while for the 
complex fields the related quantities
$$
w \propto \vec {\cal E}^2 + \vec {\cal H}^2 , ~~~~~~\vec p \propto \vec {\cal 
E} 
\times \vec {\cal H} 
\eqno(23)
$$
vanish in account of the self-duality conditions (20). Some preferance of the 
$\Re$-part fields $\vec E, \vec H$ may be therewith justified from geometrical 
considerations (see Section 5).

In addition to all this it may be shown ~\cite{GravCosm} that the structure 
of the UEqs, and of the $B$-connection (9) in particular, make it possible 
to define a $C$-valued {\it Yang-Mills field}. Infact, for the connection (9) 
one obtains the expression
$$
\Gamma_\nu = A(X)*\sigma_\nu = A^\mu(x)\sigma_\mu*\sigma_\nu = A^\mu(x) B^\rho
_{\mu\nu}\sigma_\rho \equiv A_\nu(x) + N^a_\nu(x)\sigma_a ,
\eqno(24)
$$
$B^\rho_{\mu\nu}$ being {\it the structure constants} of $B$-
algebra. Then the {\it trace-free-part} variables $N^a_\nu(x)$,
$$
N^a_o = A_a(x); \qquad N^a_b = \delta_{ab} A_o(x) + i\varepsilon_{abc}A_c(x)
\eqno(25)
$$
may be identified with the potentials of some Yang-Mills (YM) field of a 
special type.\footnote{The YM potentials like (25) belong to the class of 
so called {\it invariant} fields}

The trace-free part of $B$-curvature tensor (15) gives then for the YM field 
strengths
$$
{\cal L}^a_{[\mu\nu]} = \partial_{[\mu}N^a_{\nu]} - \frac{i}{2}\varepsilon_{abc}
N^b_\mu~N^c_\nu .
\eqno(26)
$$
For a nonzero solution $\xi(x)$ it follows now from eqs.(14) {\it for each of 
$[\mu\nu]$ component}
$$
\det \Vert R_{[\mu\nu]} \Vert \equiv {\cal F}^2_{[\mu\nu]} - {\cal 
L}^a_{[\mu\nu]}
{\cal L}^a_{[\mu\nu]} = 0,
\eqno(27)
$$

In view of (27) {\bf EM field (21) should be regarded as a modulus of 
isotopic vector of YM-triplet field}, both fields being 
described through a unique $B$-connection (9) and the EM field being 
correspondent to the trace part of it.

Such an interrelation between EM and YM fields proposed firstly in ~\cite
{GravCosm} is gauge invariant and requires no participation of a chiral 
field as in usual gauge approach. 
However, the subset of YM fields (25) can't 
be {\it pure}, being always accompanied by an EM field, due to the 
definite norm of the isotopic 3-space (see (27)).

All of the above speculations would be significant if only the YM eqs. 
would be satisfied for the field variables (25),(26). Fortunately, it is just 
the case, since the trace-free part of the self dual $B$-curvature (15) 
includes only the corresponding self-dual configuration of Maxwell strength 
tensor and the Lorentz inhomogeneous form, both being null in account of the 
integrability conditions (17),(18). Thus, for each solution to the UEqs (8) 
{\bf the YM field strengths (26) are self-dual and satisfy therefore the YM 
eqs. for free space}.

It may be noted in conclusion that, contrary to the EM case, the $\Re$ and 
$\Im$ parts of the $C$-valued strengths (26) won't satisfy YM eqs. 
separately; so the YM fields considered are essensially complex in nature! 
On the other hand, it may be checked that the YM strengths 
(2) preserve non-Abelian (commutator) part for the potentials of the form (25).
\footnote{As to the possible nonAbelian nature of EM field 
itself, it was discussed recently in ~\cite{Pestov}, also in the framework 
of Weyl geometry}

\bigskip
\noindent
{\bf 5.Dion-like unisingular solution and quantization of electric charge}

The vacuum Maxwell eqs. hold identically for each solution to the generating 
UEqs. Hence, no soliton-like field distributions can exist for the model 
considered. Nevertheless, {\bf the particles may be brought into 
correspondence with the singular points (or strings, membranes etc.) of the 
field functions}, in which the $B$-differentiability conditions are violated.

In account of the self-dual structure  of  gauge  fields  proved, 
{\bf charged singular solutions}, if exists, {\bf should  be 
dions}, i.e. carry both the electric and magnetic charges  of 
equal (up to a factor "i") magnitudes!
Elementary unisingular  dion-like  solution  has  been  found  in 
~\cite{AlgDyn}. To obtain it here, we can fix the gauge  so  that 
to have for the 2-spinor $\xi(x)$ the form
$$
\xi^T(x) = (~1,~G(x)~) ;
\eqno(28)
$$
then for the EM potentials one gets
$$
A_w = \partial_u G,~~~A_v  =  \partial_{\bar  w}  G,  ~~~  A_u  = 
A_{\bar w} \equiv 0,
\eqno(29)
$$
and the system of UEqs (8) is reduced to a couple of nonlinear 
differential eqs. for a unique unknown function $G(x)$ only
$$
\partial_w  G  =  G  \partial_u  G,  \qquad  \partial_v  G  =   G 
\partial_{\bar w} G,
\eqno(30)
$$
where the {\it light-cone  coordinates}  $\{u,v,w,\bar w\}$  defined 
previously by eq.(5) have been used.

By  mutual  multiplication  of  eqs.(31)  we  come  then  to  the 
connection
$$
(\partial_u G)(\partial_v G) - (\partial_w G)(\partial_{\bar w}  G) 
= 0,
\eqno(31)
$$
which  appears  to  be  just  the  eikonal  eq.(7)  in   spinor 
coordinates. Now we obtain the commutator of derivatives in the left 
side of eqs.(30) and get, in respect to eq.(31),
$$
\partial_u \partial_v G - \partial_w \partial_{\bar w} G = 0,
\eqno(32)
$$
the latter being just the wave d'Alembert equation $\bigtriangledown^2 G =  0$. 
It may be shown that this result is gauge invariant, so that {\bf 
each component of the 2-spinor field  $\{\xi_B(x)\}$  obeys  both 
the eikonal and d'Alembert eqs. (31),(32)!}

Fundamental non-trivial solution common to eqs.(31),(32) found in 
~\cite{AlgDyn}, corresponds to the {\it stereographic 
mapping\/} $S^2 \mapsto C$ of the 2-sphere onto the complex plane:
$$
G = \frac{x^1 + ix^2}{x^3 \pm r} \equiv \frac{\bar w}{z \pm r} \equiv 
\tan^{\pm 1}\frac{\theta}{2} \exp^{i\varphi},
\eqno(33)
$$
$\{r,\theta,\varphi\}$ being usual spherical coordinates. From the solution 
(33), which satisfy the couple of eqs.(30) under consideration, using the 
expressions (29) the EM potentials $A_w, A_v$ may be found; then for the 
scalar ($A_o$) and spherical components we get
$$
A_o = \pm\frac{1}{4 r},~~A_r = -\frac{1}{4 r}, ~~A_\varphi = \pm iA_\theta = 
\frac{i}{4 r} \tan^{\pm 1}\frac{\theta}{2}
\eqno(34)
$$
Now for the nonzero components of $C$-valued EM field strengths (21) we get 
(the electric field appears to be pure real, while the magnetic - pure imaginary!)
$$
{\cal E}_r = \pm \frac{1}{4 r^2}, ~~~ {\cal H}_r = \pm \frac{i}{4 r^2},
\eqno(35)
$$
(note that the components $A_r,~A_\theta$ don't contribute into the magnitude 
of field strengths, being of a pure gauge type).
Hence, the fundamental unisingular solution (33) corresponds to {\bf a 
point source with a fixed value of electric charge $q = \pm 1/4$ and an 
equal (imaginary) value of magnetic charge $m = \pm i/4$}.

At this stage of consideration, the factor $(1/4)$ isn't of great importance, 
since the {\it physical} EM potentials were determined up to an 
arbitrary dimensional factor only. What is really 
significant is that 1) all values of electric charge except {\it the only 
possible one} are not allowed for the point particle-like source to possess,
and 2) its Coulomb field is always accompanied by the magnetic monopole 
field with the charge equal to the electric one! 

We set aside the general problem of charge quantization in this and 
similar (see ~\cite{CCF}) models in hope to discuss it elsewhere, and 
consider now an interesting modification of the solution (33)-(35), 
which can be obtained through the {\it complex translation\/} $z \mapsto 
z + i a, ~~a \in R$, the latter being obviously a symmetry of UEqs. By this 
we come to a new solution, whose electric field structure instead of Coulomb 
form (35) corresponds to a known Appel solution (see e.g. ~\cite 
{Newman}). The singular EM -field structure will be then defined by the 
condition
$$
r^* \equiv \sqrt{(z+ia)^2 + x^2 + y^2} = 0,~~\Rightarrow~~\{ x^2 + y^2 = a^2, ~~
z = 0\},
\eqno(36)
$$
corresponding to a {\it ring-like} source of radius $|a|$. For the real-part 
fields ($\Re$-fields) $\{\vec E, \vec H\}$ the following asipmtotic expressions 
at the distances $r >> |a|$ are true:
$$
E_r \simeq \frac{q}{r^2} (1 -\frac{3 a^2}{2 r^2}(3\cos^2{\theta}-1)), 
~E_\theta \simeq -\frac{q a^2}{r^4} 3\cos{\theta}\sin{\theta}, 
~H_r \simeq \frac{2 q a}{r^3}\cos{\theta}, ~H_\theta \simeq \frac{q a}{ r^3}
\sin{\theta}.
\eqno(37)
$$
In view of eqs.(36)-(37), the $\Re$-field solution is related to the 
singular ring with a quantized value of electric charge $q = \pm 1/4$, {\it 
dipole magnetic moment\/} $\mu = qa$ and {\it quadrupole electric moment\/} 
$\vartheta = -2qa^2$. If we assume now for $|a|$ the value $|a|=\hbar/2Mc$ ~in 
order to have for the magnetic moment the known Dirac value $\mu=e\hbar/2Mc$, 
~then it may be conjectured {\bf for a fundamental charged fermion to possess 
necessarily a quadrupole electric momemt} $\vartheta$ equal in magnitude to 
$$
\vartheta = \frac{e \hbar^2}{2 M^2 c^2} 
\eqno(38)
$$
At present, such a statement looks rather speculative; nontheless, 
the possibility of its experimental prove may be discussed. However, 
much more fundamental seems to be the fact that for the $\Re$-part fields 
their asimptotic structure (37) is in complete agreement with that observed 
for elementary particles, whereas the $\Im$-fields contains only the "phantom" 
terms proportional, say, to the magnetic charge or to the dipole electric 
moment!

Geometrically, the phantom-like $\Im$-fields contribute only into the 
{\it torsion} terms of the "Minkowsky-projection" of the complex 
$B$-connection (9). Then, owing to a specific 
(totally skew symmetric {\it Rodichev-type}) structure of the torsion considered, 
the $\Im$-fields won't enter into the eqs. of {\it geodesics}, this resulting in 
{\bf total non-observability of magnetic charges and electric dipole moments 
for the elementary particles} (see ~\cite{GravCosm} for details). 

\bigskip
\noindent
{\bf 6.Effective metric, twistor variables and implicit general solution 
to UEqs}

In a special gauge the UEqs were shown to reduce themselves to the couple of 
eqs.(30). The latter is known in GTR as the system defining a {\it shear-free 
geodesic null congruence} (SFGNC)~~$l_\mu(x)$, for which we can take, say, 
the spinor representation
$$
l_\mu \Leftrightarrow L = {1 ~~~\bar G \choose G ~~\bar G G} , ~~~ \det\Vert L 
\Vert \equiv 0.
\eqno(39)
$$
Then the induced  Riemannian metric of a {\it Kerr-
Schild type}
$$
ds^2 = du dv - dw d\bar w - M\Re(\partial_{\bar w} G)(du + Gd\bar w + \bar G dw
+ G\bar G dv)^2 / (1 + G\bar G)
\eqno(40)
$$
in stationary case satisfy the vacuum Einstein equations ~\cite{KerrDS, Burin}. For unisingular 
solution above-presented the metric (40) 
{\bf appear to be just of Schwarzschild (for 
the point singularity (33)) or Kerr (for the ring singularity (36)) types!}

We pass now to the demonstration of complete integrability of the system (30)
based on a famous Kerr theorem (see~\cite[chapter 7]{Penrose}). In respect to it, the 
general solution to the SFGNC system of eqs.(30) has the form of implicit 
dependence of $G(x)$ 
$$
F(G, \tau_1, \tau_2) = 0
\eqno(41)
$$
upon the (projective) {\it twistor variables}
$$
\tau = X\eta \qquad (\tau_1 = u + w G, ~~~\tau_2 = \bar w + v G ).
\eqno(42)
$$
The {\it caustic} condition
$$
\frac{d F}{d G} = 0
\eqno(43)
$$
is ten known to define the singular points (or strings etc.) of the curvature 
of the Kerr-Schild metric (40). On the other hand, {\bf the same condition 
(43) may be verified to fix the singularities of EM-field} which can be 
constructed from the $G(x)$-function ~through the 4-potentials (29). 

Thus, we reduced the UEqs (8) to the SFGNC-system (30) and the latter - to a 
purely algebraic problem of resolving of the implicit-functional dependence 
(41). The last problem, rather complicated in its turn, will be discussed 
elsewhere.

\bigskip
\noindent
{\bf 6.Perestroikas of singularities as mutual transmutations of particles}

The structure of UEqs (1), purely algebraic in origin and compact in form, 
appears to be very complicated, being related to spinors, twistors and gauge 
fields. Each solution to (1) satisfy, in particular, Maxwell eqs., whose 
singular structure is found to be quite nontrivial (see below) and may be 
therefore identified with that of elementary particles. 

On the other hand, not an every solution to the linear Maxwell eqs. corresponds 
to some any solution to the primary system (1), from where the quantization 
of electric charge does follow, as well as the nontrivial time evolution of 
the particles-singularities.

From a purely mathematical point of view, all this bears a direct relationship 
with the rapidly progressing {\it catastrophe theory} ~\cite{Arnold}, in 
whose framework {\bf the "perestroikas" of singularities should be interpreted 
as the processes of mutual transmutation of elementary particles}.

The confirmation for such a conjecture we have ontained recently ~\cite{Abstr}, 
where an exact bisingular solution to the UEqs (and, hence, to the ordinary 
Maxwell eqs. as well) was presented. Its structure describes the axial-symmetric 
interaction of two point-like oppositely charged singularities, the magnitudes 
of charges being equal to the charge of unisingular solution (35). Under some 
values of integration constant, this solution describes also the {\it 
"creation - annihilation" processes} of particles - singularities, together with 
{\it an intermediate resonance structure of a toroidal geometry}. We hope to 
examine it in detail in the near future, as well as the general problem of 
interactions of singularities in the framework of the UEqs model, which may 
give rise to many other striking phenomena.

One of the authors (V.K.) is grateful to the participants of the conference for 
helpful discussions, and to the organizers - for hospitality in Kazan.


\bigskip
\begin{thebibliography}{99}
\bibitem{AlgDyn}
Kassandrov V.V. Algebraic structure of space-time and algebrodynamics. - 
Moscow, Peop. Friend. Univ. Press, 1992 (in Russian).
\bibitem{CCF}
Kassandrov V.V., Rizcalla J.A. Covariantly constant fields and geometrization 
of electromagnetism. // Proc. Int. Conf. "Geometrization of physics II". Kazan, 
Kazan Univ. Press, 1996. -- p.137 (in Russian).
\bibitem{Shir}
Vishnewsky V.V., Shirokov A.P., Shurygin V.V. Spaces over algebras. Kazan, 
Kazan Univ. Press, 1985 (in Russian).
\bibitem
{Chrenn}
Khrennikov A.Yu. Superanalysis. Moscow, "Nauka", 1997 (in Russian). 
\bibitem{Depon}
Kassandrov V.V. Analytic superfunctions and dynamics of physical fields. Moscow, 
VINITI Ac. Sci. USSR, No.152-80 DEP, 1980 (in Russian).
\bibitem{Klu}
Kassandrov V.V. Conformal mappings, hyper-analiticity and field dynamics. // 
Acta Applic. Math., 1997 (to appear); Physical fields as superanalytic 
mappings on the algebraic structure of space-time. // In "Quazigroups and 
nonassociative algebras in physics". Tartu, Proc. Inst. Phys. Estonia , No.66, 
1990. -- p.202.
\bibitem{GravCosm}
Kassandrov V.V. Biquaternionic electrodynamics and Weyl-Cartan geometry of 
space-time. // Grav. \& Cosm., {\bf 1}, No.3, 1995. -- p.216.
\bibitem{Westnik}
Kassandrov V.V. Electromagnetic waves as hyperanalytic mappings. // Vestnik 
Peop. Friend. Univ., Fizika, No.1, 1993. -- p.59 (in Russian).
\bibitem{Gursey}
Evans M., G$\ddot u$rsey F., Ogievetsky V. From two-dimensional conformal to 
four - dimensional self-dual theories: Quaternionic analyticity.// Phys.Rev.D, 
{\bf 47}, 1993. -- p.3496.
\bibitem{Kretch}
Krechet V.G. Geometrization of physical interactions, 5-dimensional theories 
and the many-world problem. // Grav. \& Cosm., {\bf 1}, No.3, 1995. -- p.199.
\bibitem{Pestov}
Barbashov B.M., Pestov A.B. Weyl connection and nonAbelian gauge group. // 
Theor.Math.Phys., {\bf 104}, 1995. -- p.429 (in Russian). 
\bibitem{Newman}
Newman E.T. Maxwell equations and complex Minkowsky space. // J.Math.Phys., 
{\bf 14}, 1973. -- p.102.
\bibitem{KerrDS}
Debney G., Kerr R.P., Schild A. Solutions of the Einstein and Einstein-Maxwell 
equations. // J.Math.Phys., {\bf 10}, 1969. -- p.1842.
\bibitem{Penrose}
Penrose R., Rindler W. Spinors and space-time. II. Cambridge Univ. Press, 1986.
\bibitem{Burin}
Burinskii A.Ya. String-like structures in complex Kerr geometry. // Rep. on 
IV Hungar. Relat. Workshop, G\'ardony, 1992.
\bibitem{Arnold}
Arnold V.I. Singularities of caustics and wavefronts. Moscow, "Matematica", 1996.
\bibitem{Abstr}
Rizcalla J.A., Kassandrov V.V. Bisingular solutions in biquaternionic 
electrodynamics. // XXIII Phys.-Math. Conf. Peop. Friend. Univ. Abstracts. 
Moscow, 1997. -- p.97.

\end{thebibliography}
\end{document}